# Atomic and mesoscopic structure of Dy-based surface alloys on noble metals

Sina Mousavion[1], Ka Man Yu[1], Mahalingam Maniraj[1,2], Lu Lyu[1], Johannes Knippertz[1], Benjamin Stadtmüller[1,3] Martin Aeschlimann[1]

[1]Department of Physics and Research Center OPTIMAS, University of Kaiserslautern, Erwin-Schrödinger-Straße 46, 67663 Kaiserslautern, Germany
[2]Deparent of Physics, Indian Institute of Technology Bombay, Powai, Mumbai 400076, Maharastra, India
[3]Institute of Physics, Johannes Gutenberg University Mainz, Staudingerweg 7, 55128 Mainz, Germany
*Email: sina.mousavion@gmail.com, bstadtmueller@physik.uni-kl.de

**Abstract**: Surface alloys are a highly tunable class of low dimensional materials with the opportunity to tune and control the spin and charge carrier functionalities on the nanoscale. Here, we focus on the atomic and mesoscopic structural details of three distinctive binary rare-earth-noble metals (RE/NM) surface alloys by employing scanning tunneling microscopy (STM) and low energy electron diffraction (LEED). Using Dysprosium as the guest element on fcc(111) noble metal substrates, we identify the formation of non-commensurate surface alloy superstructures, which leads to homogeneous moiré patterns for $DyCu_2$/Cu(111) and $DyAu_2$/Au(111), while an inhomogeneous one is found for $DyAg_2$/Ag(111). The local structure was analyzed for these samples and the observed differences are discussed in the light of the lattice mismatches of the alloy layer with respect to the underlying substrate. For the particularly intriguing case of a $DyAg_2$ surface alloy, the surface alloy layer does not show a uniform long-range periodic structure, but consists of local hexagonal tiles separated by extended domain walls, which occur likely to relieve the in-plane strain within the $DyAg_2$ surface alloy layer. Our findings clearly demonstrate that surface alloying is an intriguing tool to tailor the local atomic structure as well as the mesoscopic moiré structures of metallic heterostructures.

# Introduction

The growing demand for future technological applications to transport, process and store digital information on ever smaller length scales has triggered the search for novel low dimensional materials with unconventional spin and charge carrier functionalities. Apart from intrinsically 2D materials such as atomically thin honeycomb structures [1, 2] or transition metal dichalcogenides [3, 4], binary surface alloys have become an intriguing choice for engineering ultrathin structures with tunable material properties. Surface alloys can be fabricated with long-range order and high structural quality on noble metal surfaces by replacing surface atoms of the noble metal host materials with different kinds of metallic guest atoms. This allows the investigation of bonding mechanisms for otherwise immiscible metals and to design the spin-dependent band dispersion of surfaces states [5–11].

The most frequently studied class of surface alloys is heavy metal noble metal (NM) surface alloys grown on fcc(111) noble metal surfaces. Typical examples for this class of surface alloys are BiAg$_2$, PbAg$_2$, Bi$_x$Pb$_{1-x}$ [12–15] surface alloys which all exhibit a giant Rashba split surface state in the vicinity of the Fermi energy. Crucially, the spin-splitting, as well as the energetic position of this Rashba-type surface state, strongly depends on the spin-orbit coupling strength of the guest and host materials as well as on the vertical surface relaxation of the guest atoms [16–22]. This offers an intriguing opportunity to control the spin-split band structure of the material at the Fermi energy and hence to tailor the corresponding spin functionalities of these materials for spintronic applications.

Recently, the observation of ferromagnetism in a rare-earth (RE)-based surface alloy with high Curie temperature [23] and topological characteristic [23–26] have revived the research on surface alloys. Interestingly, RE-based surface alloys exhibit intriguing structural properties that are clearly distinct from those of heavy metal-based surface alloys. All heavy-metal noble-metal surface alloys on fcc(111) substrates form the same $\sqrt{3} \times \sqrt{3}$ R30° superstructure in which one guest atom per unit cell always replaces one host atom of the surface layer. This results in an equal density of atoms in the first and second layers of the material. The situation is significantly more complex for RE-based surface alloys for which the density of atoms in the alloy layer (first layer) and the second layer are typically different. This results in the formation of moiré patterns for RE/NM surface alloys depending on the element of the guest and host atoms. It is established that the density of states of the alloy layer is periodically modulated within the moiré superstructure [26]. An interesting consequence of this complex pattern is the moiré-driven hybridization of electronic states near the Fermi level for various RE/NM surface alloys [24] that can arise due to the lattice strain within the alloy layer as well as the partial vertical buckling of the guest atoms [24, 27]. For instance, Correa et al. observed electronic states for the GdAg$_2$ surface alloy that depend on the parameters of the moiré lattice [27]. In addition to the atomic superstructures, recent studies reported the existence of long-range ordered moiré patterns for the GdAg$_2$/Ag(111) and GdAu$_2$/Au(111) surface alloys with periodicities of the surface alloy layer of nearly (12 × 12) on (13 × 13) grid of the bare Au(111) and Ag(111) substrates [23, 28]. A study of the GdAg$_2$ reports strain within the alloy layer (1-7 %) that can be tuned by changing the substrate's temperature during the Gd deposition and further a tessellation of the alloy layer [27]. CeAu$_2$/Au(111) and LaAu$_2$/Au(111) exhibit lattice periodicities of (10.6 × 10.6) of the CeAu$_2$ surface alloy on a (11.6 × 11.6) grid of the bare Au(111) surface, and of a (10.4 × 10.4) LaAu$_2$ lattice on a (11.4 × 11.4) Au(111) grid [24]. Similarly, ErCu$_2$/Cu(111) has a lattice periodicity of (8.7 × 8.7) ErCu$_2$ on (9.6 × 9.6) Cu(111) [29]. Despite these numerous investigations of a variety of RE/NM surface alloys [5, 23, 24, 28–30], there is still no clear model that can predict the observed superstructures based on the chemical composition and surface orientation of the constituents of the surface alloy. In this regard, these characteristic structural (superstructure and moiré pattern) and corresponding electronic properties make RE-based surface alloys a compelling substrate for potentially functionalizing the properties of organic and inorganic adsorbate structures grown on top [23, 31–35].

In the present work, we, therefore, focus on a comprehensive study of the atomic and mesoscopic structure of three surface alloys formed between the guest atom Dy and the (111) surfaces of the noble metals Cu, Ag, and Au. Our results reveal identical stoichiometry for all three systems, i.e. a 2:1 ratio between the host and the guest element, Dy. Each alloy forms a unique non-commensurate superstructure with different degrees of complexity. As a result of their atomic unit cells, all three surface alloys exhibit additional long-range orders in the form of complex moiré patterns due to the atomic mismatch between the surface alloy and its respective substrate. The DyAu$_2$ exhibits a relatively straightforward periodic structure. In the case of DyCu$_2$, two mirror domains are developed and a highly periodic structure is maintained. In the case of DyAg$_2$, we find a non-uniform moiré pattern with complex electron density modulation in the STM, which reveals a strain-relief mechanism in the form of surface tessellation. Despite an almost identical interatomic distance of Au(111) and Ag(111), we observe significantly different surface structures.

## Experimental Details

**Sample preparation.** All samples were prepared in an ultra-high-vacuum (UHV) preparation chamber with a base pressure $< 5 \times 10^{-10}\ mbar$. This preparation chamber is equipped with low-energy electron diffraction (LEED) optics and a Dy evaporator.

The noble metal surfaces were prepared by Ar$^+$ sputtering using energy ranges between $1.2 - 1.5\ keV$ for 30 min and subsequent annealing at $T_{Cu} = 845\ K, T_{Ag} = T_{Au} = 860\ K$ for 30 – 45 min. The cleanliness of the surface was confirmed by a low defect density of the surface in a large scale room temperature scanning tunneling microscopy (STM) experiment and the narrow spot profile of the three-fold diffraction spots of the fcc(111) surfaces acquired by LEED for energies between $50 - 120\ eV$.

Dy was evaporated from a tungsten crucible using a FEM-3 Focus e-beam evaporator at a constant flux that was monitored using the flux monitor controller. The substrates were kept at constant temperatures ($\sim 570 - 630\ K$) during the Dy deposition (base pressure $< 5 \times 10^{-10}\ mbar$). The sample temperatures are measured using a K-type thermocouple in contact with the sample plate. The freshly prepared surface alloys were held at ~20 degrees lower temperatures for a few minutes after the deposition to improve surface quality.

**STM.** The STM measurements were performed in a UHV system directly connected to the preparation chamber (VT-AFM XA, Omicron GmbH, with base pressure$< 2 \times 10^{-11}\ mbar$). All measurements were carried out in constant current mode. In our setup, the biased voltage is applied to the tip. The sample temperature was kept at 106 K using a liquid $N_2$ cooling system. Tungsten tips were electrochemically etched, and in-situ cleaned by $Ar^+$ Sputtering ($1.5 - 4.0\ keV$) and subsequently annealed using a direct current to reduce impurities and achieve an atomically sharp apex. The STM images were processed and analyzed using WSxM [36] and Gwydion [37] software.

**LEED**. LEED measurements were performed using a rearview four grid system from Scienta Omicron at room temperature. Images were acquired using a CCD camera attached directly at the viewport of the LEED optics. To visually enhance the image contrast, all images were post-processed using the ImageJ software. The simulation of the diffraction pattern was performed using the Spot-Plotter Software [38]. The $DyAu_2$ data were acquired using an MCP LEED in another UHV chamber, for which the image was corrected using LEEDCal [39]. The error values of the superstructure matrixes are estimated by fitting a range of simulated diffraction patterns that deviate from the best-fit values. This is done by changing the radial and azimuthal parameters limited by the full width at half maximum of the spot profile. We then used the largest deviation value. To obtain the error bars for the real space unit cell parameters, LEED images are evaluated by computing the positions of the surface alloy and the substrate diffraction spots, which are obtained by profile fitting of the 2D Gaussian function. Error propagation of these quantities is based on the half width at half maximum of the fitted spot profile.

## Results

In this section, we will present LEED and STM results for the surface alloy superstructures of Dy on Au(111), Cu(111), and finally, Ag(111).

## DyAu2/Au(111)

### Periodic order of the DyAu$_2$/Au(111) surface alloy

Fig 1(a) shows the LEED pattern of the DyAu$_2$/Au(111) surface recorded for electron energy of 65 eV. Besides the substrate spots (blue), the LEED pattern reveals a set of diffraction spots that point to the formation of a higher-order DyAu$_2$ surface alloy superstructure. The best agreement between the experimental data and the simulated LEED pattern was obtained for the structure with the superstructure matrix $\begin{bmatrix} 2.17 \pm 0.04 & 1.09 \pm 0.04 \\ -1.09 \pm 0.04 & 1.09 \pm 0.04 \end{bmatrix}$. The top left quarter of the LEED data is superimposed with the first-order diffraction spot of the substrate (blue) as well as the first and second-order spots of the alloy (yellow). The corresponding alloy superstructure exhibits a hexagonal unit cell with a lattice constant of ~5.41 Å (~1.88 × $d_{Au}$, where $d_{Au} = 2.88$ Å) in real space which is rotated 30° with respect to the <$\bar{1}10$> direction of the substrate. In addition, we observe a second diffraction pattern with a large number of diffraction spots with varying intensity. This diffraction pattern could be modelled by the superstructure matrix $\begin{bmatrix} 13 \pm 0.27 & 0 \pm 0.27 \\ 0 \pm 0.27 & 13 \pm 0.27 \end{bmatrix}$. Within the experimental uncertainty, this superstructure revels a close to commensurate or even a commensurate registry with respect to the surface grid. The corresponding diffraction spots are superimposed onto the experimental diffraction image as red circles and the unit cell in k-space is marked by a green rhombus. This second diffraction pattern with its large ~37.44 Å unit cell in real space is assigned to the moiré pattern of the DyAu$_2$/Au(111). The existence of such a moiré already points to an atomic mismatch between the surface alloys layer and the topmost layer of the Au(111) substrate. More details about the moiré pattern will be discussed in the following section.

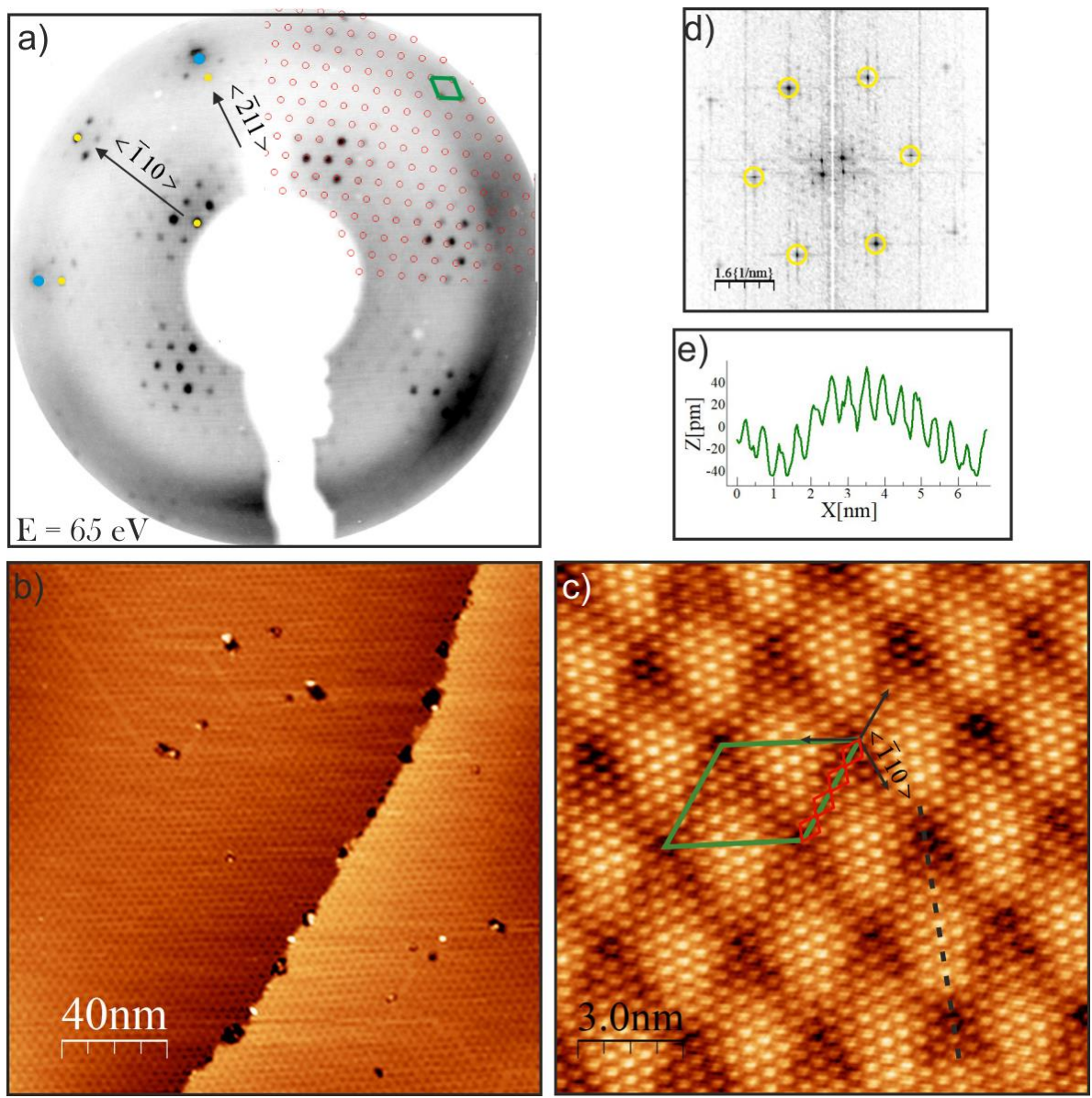

**Fig 1. DyAu₂/Au(111) surface alloy (a) LEED measurement at E = 65 eV, (b) STM measurement of (b) 200 × 200 nm² scanned region including uniformly grown superstructure at 270 K ($I_{set}$= 200 pA, $V_{Bias}$= 200 mV) (c) 15 × 15 nm² area of a single domain ($I_{set}$= 270 pA, $V_{Bias}$= -350 mV). Black <01$\bar{1}$> vectors are the substrate orientations. (d) FFT of (c). (e) Line scan along the black dashed line between two moiré spots in (c)**

## Surface structure of DyAu$_2$/Au(111)

Next, we studied the DyAu$_2$ surface alloy structure by STM to understand the local atomic and mesoscopic structure in the real space. The large area in Fig 1(b) shows that the herringbone reconstruction of the Au substrate is replaced by uniform moiré patterns. This is an indication that the monolayer surface alloy forms successfully, similar to the earlier report for the GdAu$_2$ surface alloy [25]. Fig 1(b) shows two adjacent terraces with uniformly grown DyAu$_2$ surface alloy structures. A more detailed view of the structural properties of the DyAu$_2$ surface alloy can be obtained in the atomically resolved STM image in Fig 1(c). We observe periodically ordered bead-like features, which are superimposed on the moiré pattern-induced contrast variation. These bright protrusions are separated ~5.5 Å. They can be attributed to Dy atoms with enhanced electronic contribution to the tunneling process, similar to earlier works [23–25, 28]. Thus, we assume and assign the small bright protrusions to Dy atoms which hence reflect the atomic periodicity of the DyAu$_2$ superstructure. Fast Fourier transform (FFT) of this image in Fig 1(d) reveals the periodic ordering of these protrusions in reciprocal space (marked by yellow circles), agreeing well with the diffraction pattern found in LEED. The broad dark features, on the other hand, reflect the periodicity of the moiré pattern and are caused by the reduced electron density in the moiré potential modulation. The proposed surface alloy unit cell (red rhombus) and the moiré unit cell (green rhombus) as well as the $<01\bar{1}>$ crystal orientations (black vectors) are marked in Fig 1(c). The surface alloy lattice vectors have average length of $5.5 \pm 0.3$ Å and are oriented along the substrate's high symmetry $<11\bar{2}>$ directions. The average length of the moiré lattice vectors is $\sim 36.0 - 37.0$ Å and they are oriented along the $<01\bar{1}>$ directions of Au(111). These values are in quite good agreement with our LEED analysis. A line scan along two moiré minima (dashed line in Fig 1(c)) shown in Fig 1(e) reveals a maximum apparent vertical corrugation of ~0.55 Å within the moiré unit cell. The lattice constant of both the surface alloy and the moiré are very similar to GdAu$_2$ surface alloy [28]. In both cases, the observation of the moiré pattern points out to a different density of atoms in the first and second layers of the crystal. Correa et al. reported observation of different contrast moiré patterns in GdAu$_2$ [26]. Their theoretical results reveal the origin of the contrast to be various degrees of vertical relaxation of the alloy layer from the supporting substrate to form lattice-matched periodicity. Further, they assign TOP and FCC/HCP sites to valleys and hills of the moiré pattern contrast. Their theory is further supplemented with experimental dI/dV mapping and spectroscopy which makes clear support of two types of electronic states, characteristic to each site. It should be noted that the tip/sample distance could also influence the intensity of the observed moiré contrast [40], making it difficult to identify the host metal elements (in this case Au atoms) of the surface alloy. Careful examination reveals that the minima of the moiré pattern in the STM image exhibit different arrangements of Dy atoms. These local variations suggest a non-commensurate registry between the moiré pattern and the substrate grid. This registry can either be classified as incommensurate or as higher order commensurate with a unit cell size larger than the STM image. This observation suggests a strong variation in the local atomic registry between the alloy layer and the supporting topmost layer of the gold substrate. Considering the size of the moiré lattice from our STM and LEED results we conclude that nearly 12 atoms of the surface layer (4 Dy and 8 Au atomic units) are spanned along almost 13 Au atoms of the substrate's top layer. This causes an atomic mismatch of almost 7.7 % between the layers. The irrational ratio of the atomic mismatch results in continuously altering adsorption sites within the moiré unit cell which ultimately is responsible for the varying STM contrast within the moiré unit cell in Fig 1(c). Similar incommensurate moiré patterns were also previously observed for weakly bounded 2D materials on metal surface [41, 42] and hence point to a rather weak coupling between the surface alloy layer and the substrate crystal.

## $DyCu_2/Cu(111)$

### Periodic order of the $DyCu_2/Cu(111)$ surface alloy

Fig 2(a) presents the LEED pattern of $DyCu_2/Cu(111)$ surface recorded for electron energy of 75 eV. The six most intense spots near the edges of the screen (one marked blue) are the first-order spots of Cu(111) substrate. The top left quarter of the image is superimposed with the first three orders of simulated diffraction spots of the alloy superstructure (yellow and orange dots). The pairs of separated spots along the $<\bar{1}10>$ direction of the Cu(111) are the first and third-order diffraction spots of the $DyCu_2$ surface alloy structures. It is clear that the spots are symmetric along the mirror axis of the copper substrate. Moreover, the distance between each pair of diffraction spots is monotonically increased with increasing distance from the specular reflection, confirming that they are originated from mirror domains. The position of the diffraction spots corresponds to ±28° rotation of the superstructure with respect to the $<\bar{1}10>$ direction in the real space. The diffraction pattern of the surface alloy unit cell can be represented as a $\begin{bmatrix} 2.26 \pm 0.02 & 1.20 \pm 0.02 \\ -1.20 \pm 0.02 & 1.06 \pm 0.02 \end{bmatrix}$ matrix. According to this matrix, the surface alloy structure in the real space has a lattice constant of ~5.00 Å (~$1.96 \times d_{Cu}$, where $d_{Cu} = 2.55$ Å) and is rotated ±28° with respect to the $<\bar{1}10>$ direction of the Cu(111). All additional diffraction spots are assigned to the moiré pattern of the surface alloy, similar to the earlier case of $DyAu_2/Au(111)$. The best agreement between the experimental and simulated moiré diffraction pattern is obtained for the close to commensurate superstructure matrix $\begin{bmatrix} 10 \pm 0.16 & 7 \pm 0.16 \\ -7 \pm 0.16 & 3 \pm 0.16 \end{bmatrix}$. The moiré superstructure has a periodicity of ~22.13 Å in real space and is rotated ±13.2° with respect to $<11\bar{2}>$ direction of the Cu(111) substrate. This corresponds to an almost 9×9 R30 ±13.2° reconstruction. The top-right quarter of the image is superimposed with the diffraction pattern of both mirror domains of the moiré superstructures which are indicated by green and red spots.

## Surface structure of the DyCu$_2$/Cu(111) surface alloy

Fig 2(b) presents a medium-size STM scan (20 nm × 20 nm) of a single domain of the surface alloy which simultaneously reveals signatures of both the atomic as well as of the moiré structure of the surface alloy. In analogy to the DyAu$_2$ surface alloy, we assign the bright protrusions to the Dy atoms. The larger dark spots mark the minima of the moiré pattern formed by the atomic mismatch between the alloy layer and the substrate. The periodicity of the moiré pattern can be extracted from the FFT of the STM image shown in Fig 2(d). The moiré

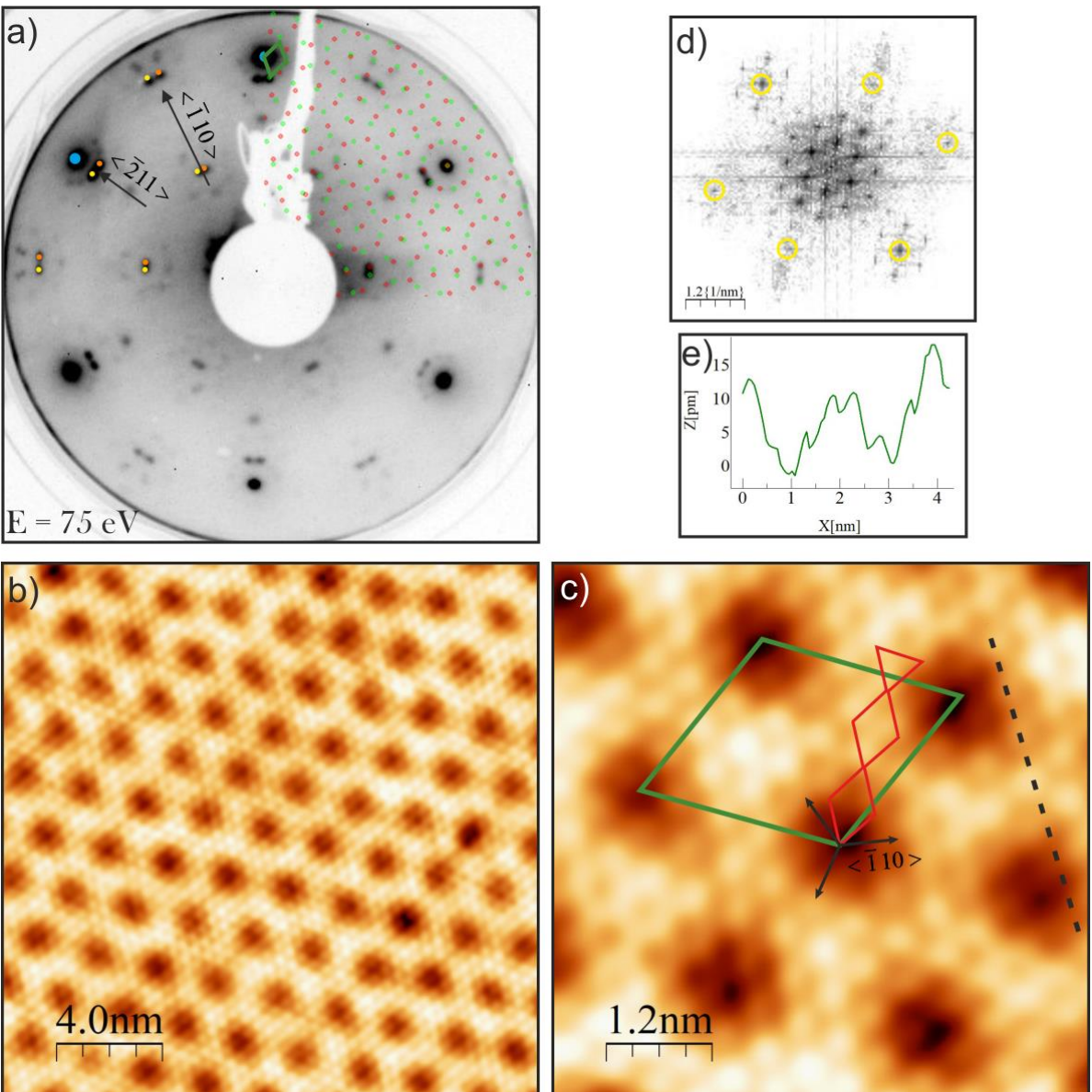

**Fig 2. DyCu$_2$/Cu(111) surface alloy: (a) LEED measurement at E = 75 eV, (b) STM 20 × 20 nm$^2$ single domain region of the surface alloy, and (c) 6 × 6 nm$^2$ selected region of the left image. Red rhombuses are the surface alloy and the large green rhombus is the moiré unit cell. Black <01$\bar{1}$> vectors are the substrate orientations. Measurements at constant current mode, I$_{set}$= 5 nA, V$_{Bias}$= 186 mV and at 106 K (d) FFT of the 15 × 15 nm$^2$ scan, (e) Line scan along the moiré spots in (c)**

unit cell clearly shows a hexagonal symmetry in the FFT pattern, in good agreement with our LEED analysis as discussed above. The central spots of the FFT and the six outer spots (circled yellow) correspond to the large moiré mesh, and the alloy superstructure respectively. A closer view of the relation between the atomic and moiré superstructure can be obtained in Fig 2(c) which presents an enlarged area of the same scan. The hexagonal arrangement of various dark depressions and bright protrusions is clearly visible in this image. The lattice formed by Dy atoms (red rhombus) is aligned by almost 29° from the $<01\bar{1}>$ direction of the Cu(111), marked by black arrows. The in-plane interatomic distance of the Dy atoms is 5.2 ± 0.3 Å indicating that the $DyCu_2$ alloy layer is successfully formed on Cu(111). This value is close to the 5.00 Å observed by LEED measurement. Importantly the interatomic distance of the atoms of the alloy superstructure is not in registry with the underlying Cu(111). As a result, nearly every 8 atoms of the alloy layer almost matches every 9 atoms of the substrate. This means an almost 11.2 % mismatch between the topmost alloy layer and the substrate, leading to the formation of a non-commensurate registry with the substrate and the corresponding moiré pattern. The latter is evidenced by the darker contrast in Fig 2(c). Similar to the $DyAu_2$ sample, the moiré minima vary in contrast and are not identical, confirming the non-commensurate nature of the alloy layer. The length of the moiré unit cell is measured 21.4 - 22.8 Å and rotated by 45.2° with respect to the $<01\bar{1}>$ direction of the Cu(111). This periodicity is in very good agreement with the proposed moiré unit cell calculated in our LEED analysis. An STM image that reveals both mirrored domains, is provided in Figure SI.4.

## $DyAg_2$/Ag(111)

### Periodic order of the $DyAg_2$/Ag(111) surface alloy

The LEED pattern of the $DyAg_2$ surface alloy, presented in Fig 3(a), is significantly different compared to the above-discussed $DyCu_2$ and $DyAu_2$. This severely complicated the unambiguous interpretation of the LEED data. For instance, the first order LEED spots of the $DyAg_2$/Ag(111) superstructure (marked as yellow points in Fig 3(a)) are split mirror symmetrically with respect to the $<01\bar{1}>$ direction of the Ag(111) substrate, similar to the $DyCu_2$/Cu(111). However, no such splitting is observed for the corresponding (11) superstructure spots along the $<11\bar{2}>$ direction (see the head of the arrow along this direction in Fig 3(a)). To resolve this apparent contradiction, we immediately turn to our local STM study to determine the structural parameters of the atomic and mesoscopic superstructure of the $DyAg_2$/Ag(111) surface alloy.

### Surface structure of the $DyAg_2$/Ag(111) surface alloy

We start our structural analysis of the $DyAg_2$ surface alloy with the overview STM scan (100 nm × 100 nm) in Fig 3(b). Similar to the $DyAu_2$ and $DyCu_2$ cases, we observe an array of dark features which we assign again to the moiré structure. However, a careful inspection reveals that these dark features cover the surface non-uniformly. In certain regions, we find a nearly uniformly-spaced grid of these features (marked by the yellow square), in others (marked red displays), the dark features are irregularly distributed. This observed non-uniformity within a single terrace is an indication that the DyAg$_2$ alloy does not grow uniformly over the Ag(111) substrate. Upon close inspection in the atomically resolved $15 \times 15\ nm^2$ STM image (Fig 3(c)), we realize that the surface consists of neighboring hexagonal tiles with surface areas varying between 15.4 - 21.8 nm$^2$ and each tile includes exactly one dark moiré valley feature. Interestingly, after scanning several different regions of the sample, there seems to be no periodic order in the distances between these tiles as well as in the tile size, However, the boundaries of the tiles are all aligned along the $<11\bar{2}>$ substrate directions. The positions of the Dy atoms appear as dark depressions in this scanned image due to the applied bias voltage. Similar to other surface alloys

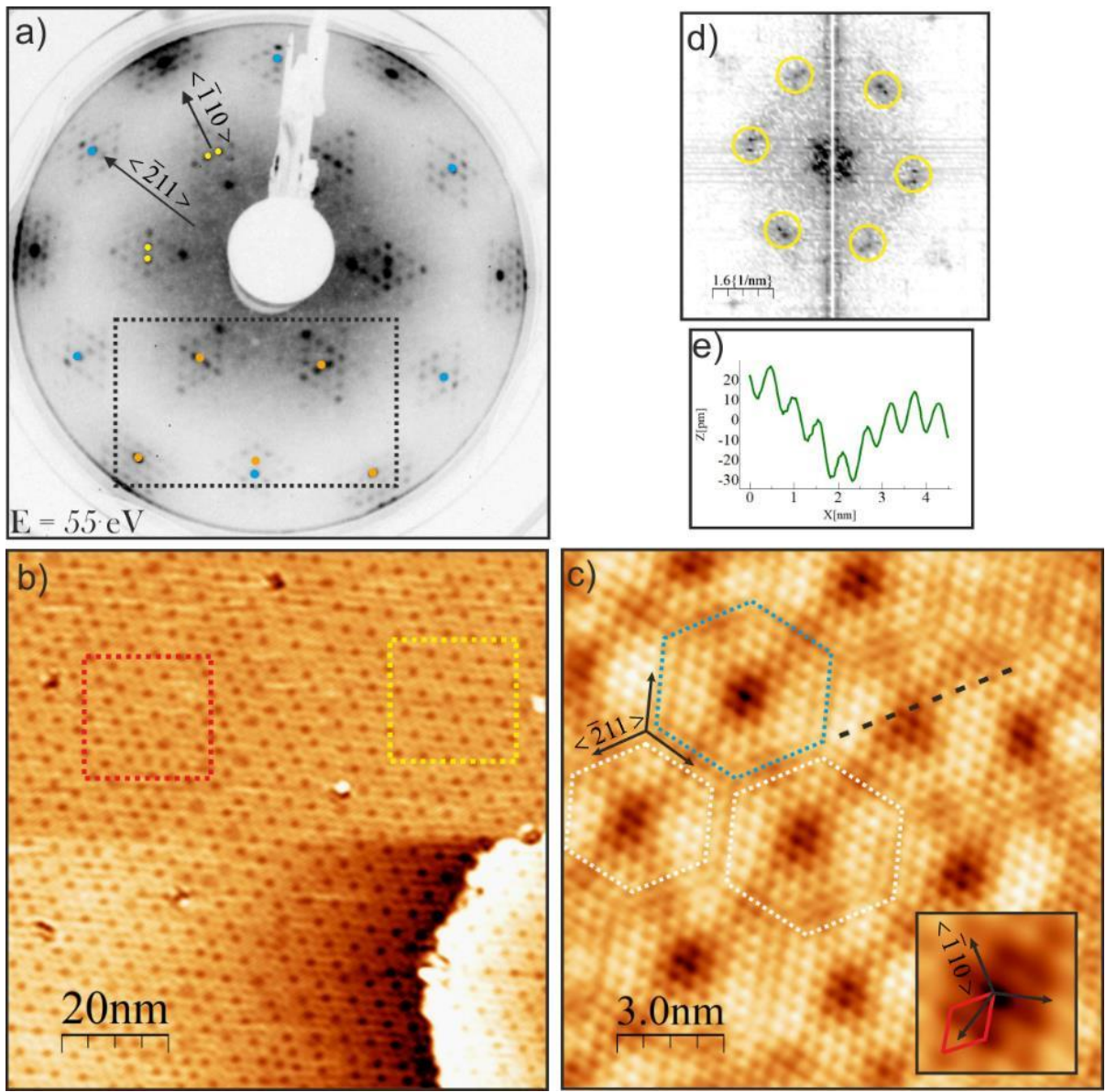

Fig 3. DyAg$_2$/Ag(111) surface alloy: (a) LEED measurement at E = 55 eV. The alloy unit cell is marked in the bottom section marked in dashed rectangle, (b) STM 100 nm × 100 nm$^2$ scanned region of the surface alloy presenting uniform (yellow square) and non-uniform (red square) features (I$_{set}$= 81 pA, V$_{bias}$= 510mV), and (c) 15 × 15 nm$^2$ scanned area. Dashed hexagons indicate domain walls of the confined structure. (I$_{set}$= 11.9 nA, V$_{bias}$= 59 mV) Inset: enlarged image of the center of the blue hexagon tile. The red rhombus is the alloy unit cells and black vectors are the $<01\bar{1}>$ substrate orientations. (d) FFT of (c). e) Line scan inside a hexagon tile in (c)

in the previous sections, the DyAg$_2$ surface alloy unit cell has a hexagonal lattice with an average unit cell vector length of 5.2 ± 0.3 Å that is rotated almost 30° with respect to the substrate's <01$\bar{1}$> orientation, creating ~3.8% atomic mismatch between the alloy layer and Ag(111) substrate. At first glance, this unit cell is in contrast to our LEED data, where we observed two sets of first-order diffraction spots (see Fig 3(a)). However, the FFT of this atomically resolved STM image (Fig 3(d)) clearly reveals double spots as well as additional intensity-modulated spots, very similar to diffraction spots observed in our LEED experiment. This point will be addressed in more detail in the discussion section. Using these structural parameters, we propose that the unit cell of the DyAg$_2$/Ag(111) surface alloy can be described by the superstructure matrix $\begin{bmatrix} 2.08 \pm 0.02 & 1.04 \pm 0.02 \\ -1.04 \pm 0.02 & 1.04 \pm 0.02 \end{bmatrix}$. The corresponding first and second-order diffraction spots are superimposed as orange points onto the experimental LEED pattern. In addition, we find that the surface alloy structure discommensurates at the edges of the tiles. As a result, the adjacent tiles are anti-phased, i.e. shifted half an alloy unit cell up or down respectively. Besides, discommensuration generates domain wall boundaries around the tiles that have rectangular lattices and an average width of 4.5 - 4.8 Å (see supplementary information (SI) 1). This distance is significantly shorter than the measured lattice size of the surface alloy unit cell and implies that the atomic units of these sites are not fully relaxed. Careful examination of the structure within the tiles confirms no additional local superstructure. Moreover, by creating a distance map of this scan, we didn't find any significant lattice distance variation within the tiles (details are provided in SI.2).

**Models and discussion**

To have a better understanding of the surface alloy structure as well as the moiré pattern formation, we present hard sphere space filling ball models of all three surface alloys in Fig 4, which are constructed based on our LEED and STM findings. We present in Table 1 the parameters of the alloy unit cell, and the moiré unit cell for each sample are extracted based on the superstructure matrixes determined experimentally. In addition, we have included calculated values for the structural parameters of the moiré unit cell based on a mathematical method for two mismatched hexagonal lattices provided by Hermann [43].

| Sample | Substrate | LEED | | | | | STM | | Hermann model | |
|---|---|---|---|---|---|---|---|---|---|---|
| | | surface alloy | | | moiré | | moiré | | moiré | |
| | $a_{sub}$ [Å] | $a_{alloy}$ [Å] | ± α [°] | | $a_{moiré}$ [Å] | ± γ [°] | $a_{moiré}$ [Å] | ±γ [°] | $a_{moiré}$ [Å] | ±γ [°] |
| DyAu$_2$ | 2.88 | 5.41(5.47±0.08) | 0 (0±0.8) | | 37.57(31±5) | 0 (0±9) | 36.0-37.0 | 0 | 37.03 | 0 |
| DyCu$_2$ | 2.55 | 5.00(4.95±0.05) | 2 (1.8±0.6) | | 22.13(25±2) | 16.8 (16±4) | 21.4-22.8 | 16.8 | 21.04 | 16.74 |
| DyAg$_2$ | 2.89 | 5.22(5.26±0.04) | 0 (0±0.5) | | NA | NA | NA | NA | 70.37 | 0 |

Table 1. Realspace lattice parameters of surface alloy superstructures, moiré unit cells, and the calculated moiré unit cell parameters using the Hermann model. For Hermann model, the mean experimental values of surface alloy parameters obtained from LEED (values outside brackets) were used. Angles α and γ are with respect to the substrate's <11$\bar{2}$> and <01$\bar{1}$> orientations respectively. The lattice parameters inside the brackets are obtained from a spot profile analysis of the experimental LEED data.

We start with the simplest structure, i.e., the one of the DyAu₂ surface alloy presented in Fig 4(a). The inset provides a structural model of the atomic unit cell of the surface alloy (red rhombus) placed on top of the Au(111) substrate (empty circles). The unit cell includes one Dy atom (green) and two Au atoms (yellow) which are placed equidistantly on the long diagonal of the rhombohedral unit cell. This stoichiometric configuration is common for all RE/NM surface alloys reported so far [23–25].

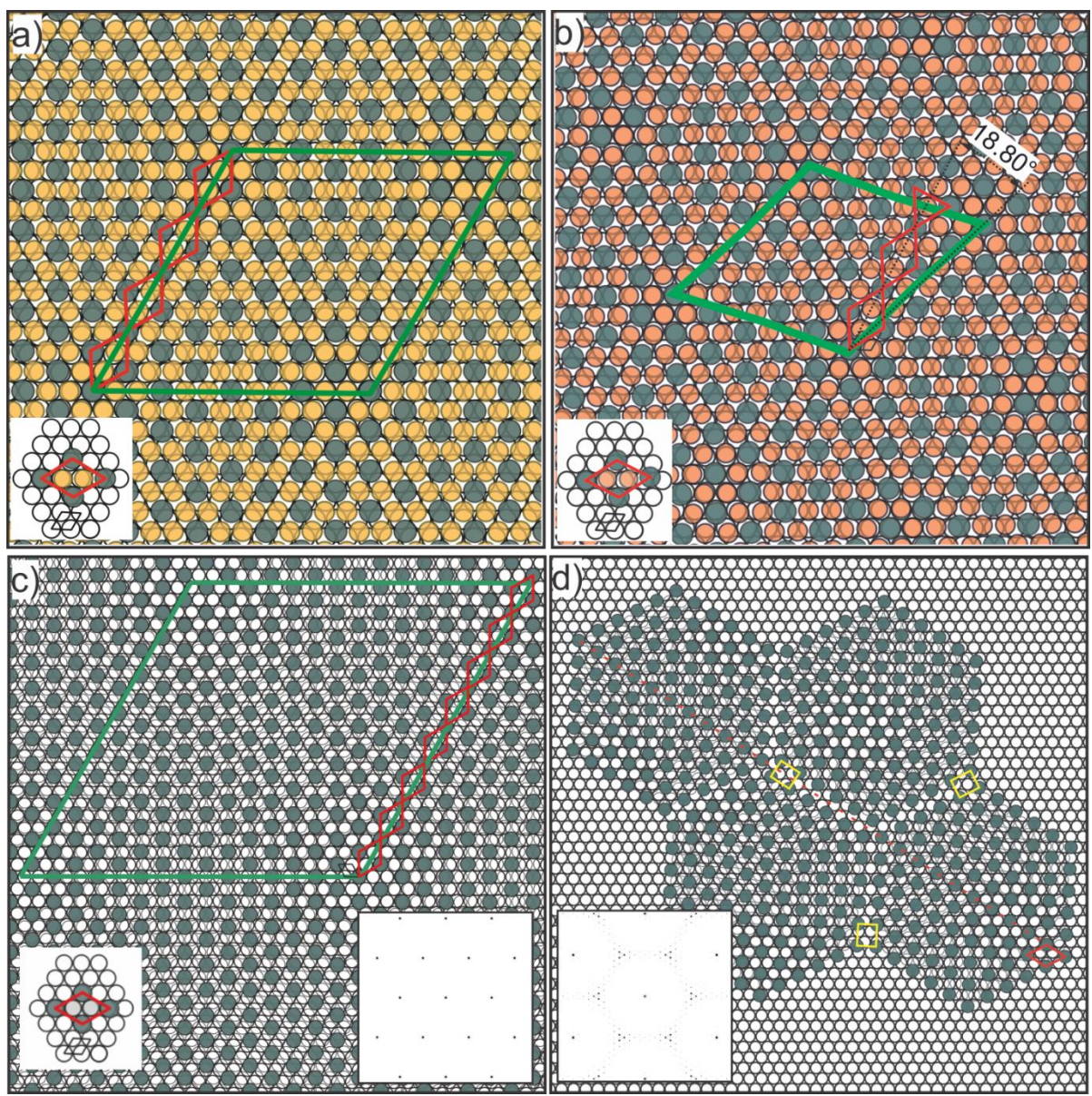

Fig 4. Ball models for (a) DyAu₂/Au(111) (b) DyCu₂/Cu(111), and (c) DyAg₂/Ag(111). Atoms of the alloy are coloured green, yellow, orange and gray for Dy, Au, Cu, and Ag respectively. The alloy and moiré unit cells are indicated using red and green rhombuses, respectively. In order to visualize the moiré pattern the top layer atoms are set with slight transparency. The inset depicts a single unit cell of the proposed surface structure on its respective bare substrate. (d) Ball model of the tessellated DyAg₂ sample based on Fig 3 (c), all 5 × 5 hexagonal tiles are identical. The Ag atoms of the overlayer are set transparent for a better visualization. The unit cell of the surface alloy and of the domain walls are marked red and yellow respectively. The red dashed line indicates the discomensuration line. The LEED simulation of a uniform DyAg₂ lattice (c) and a uniformly tessellated DyAg₂ lattice (d)

The unit cell of the DyAu$_2$ surface alloys is rotated by 30°, which is the identical orientation with respect to the substrate lattice as the typical $\sqrt{3} \times \sqrt{3}$ R30° superstructure. This structure is frequently observed for heavy metal/NM surface alloys on NM fcc(111) substrates. The only striking difference is that the interatomic distances in the DyAu$_2$ surface alloy are almost 1.08 times larger than the lattice constant of Au(111). The lattice expansion was estimated from the superstructure parameters of the superstructure analysis of our LEED data, see Table 1. This ultimately leads to the formation of a non-commensurate $\sim 1.08\sqrt{3} \times 1.08\sqrt{3}$ R30° superstructure of the DyAu$_2$ surface alloy on the Au(111) surface. Importantly, this mismatch between the atoms in the alloy layer and the substrate layer is responsible for local variation of the moiré minima in STM that are non-identical. Similarly, continuous change in the FCC/HCP-like adsorption site within the moiré unit cell generates the contrast variation that we observe inside the moiré unit cell. In terms of alloy unit cell and moiré periodicity, DyAu$_2$ surface alloy is identical to the incommensurate NaAu$_2$ surface alloy [7, 8]. In terms of alloy unit cell and moiré periodicity, DyAu$_2$ surface alloy is identical to the NaAu$_2$ and GdAu$_2$ surface alloy [7, 8, 26]. This is rather surprising since the valence electronic structure of elements are significantly different but at the same time, the atomic radii of Na, Gd, and Dy are rather similar [44]. Hence, further studies of the electronic states of different classes of surface alloys are essential to disentangle the different roles of electronic and steric effects for their structure formation of surface alloys. The ball model provides a clear visualization of the lattice mismatch between the alloy and the substrate. The moiré unit cell (Fig 4(a) green rhombus) is illustrated according to the Hermann model, based on the nearest neighbor distance of the alloy layer, and denotes its non-coinciding character. The size of this unit cell is in good agreement with our experimental result.

Next, we turn to the DyCu$_2$ surface alloy. Our LEED analysis indicates the existence of two superstructure domains mirrored along the substrate high symmetry directions, marking the DyCu$_2$ surface alloy as the only RE/NM surface alloy with 1:2 (RE:NM) stoichiometry, that is not oriented along its substrate's <11$\bar{2}$> direction. Fig 4(b) shows the corresponding atomic model structure of a single mirror domain of the DyCu$_2$ surface alloy. According to the LEED superstructure matrix analysis, the interatomic distance of the alloy layer is 2.88 Å, which is 1.13 times larger than $d_{Cu}$. The corresponding lattice expansion of the first layer is hence significantly larger for the DyCu$_2$ surface alloy compared to the DyAu$_2$ surface alloy. This distinct difference is attributed to the different atomic sizes of the Cu and Au substrate atoms, which is reflected in its lattice constant as well, refer Table 1. The mesoscopic moiré unit cell of the DyCu$_2$ alloy is rotated 18.8° with respect to the surface alloy lattice vector (marked in Fig 4(b)), which agrees both with the STM results and the calculated Hermann model. This large angle is the direct result of the ±28° rotation of the alloy layer with respect to the substrate <$\bar{1}$10> orientation, which we observe in our experimental data. As a result, the moiré features have shorter periodicity compared to the DyAu$_2$ case. Angle mismatches between the surface alloy lattice vector and moiré lattice vectors have previously been observed for RE/NM surface alloys such as LaAu$_2$/Au(111), where a smooth variation of the azimuthal ±4° misalignment of the moiré and the alloy lattice is measured and is explained as the mismatch between the alloy layer and its substrate (7.6%) [24]. However, this angle is substantially larger for the DyCu$_2$ case and is not merely a misalignment. The clear observation both in LEED and STM point out a different atomic structure of this surface alloy, compared to its related alloy family. It is possible that in addition to rotation the alloy layer minimizes its excessive energy by straining its lattice locally so that the corner Dy atoms of the moiré unit cell match the substrate lattice in the ball model. However, using only the STM result, it is not possible to directly confirm this argument.

Finally, we consider the most complex case, the DyAg$_2$ surface alloy. Considering the results of our LEED superstructure matrix analysis, the expansion of the DyAg$_2$/Ag(111) alloy layer with respect to its substrate is only 3.8 % which is significantly smaller than of the DyAu$_2$/Au(111). Fig 4(c) presents the model which is constructed based on the LEED matrix for the DyAg$_2$ surface alloy. The large green rhombus is the moiré unit cell that is based on the predicted Hermann model. Nonetheless, our STM data reveal that the DyAg$_2$ surface alloy does not follow this simple model. Instead, the structure tessellates in the form of large hexagonal tiles, which are shifted with respect to each other by half an alloy unit cell, creating anti-phase domains and boundary walls (Fig 3(c), Fig 4(d)). Based on our experimental findings, we have simulated LEED models by calculating the structure factors of uniform (Fig 4(c)) and tessellated (Fig 4(d)) DyAg$_2$ superstructures. The tessellated superstructure reveals spot splitting as well as additional intensity-modulated spots (inset of Fig 4(d) and SI.3), which indicates that the co-existence of multiple anti-phase domains caused by the surface tessellation is the genuine reason for the spot-splitting in our LEED pattern (see Fig 3(a)) [45]. Moreover, this can at least partially explain the intensity modulation that we observe in the LEED result as well. Spot-splitting has been previously reported for similar anti-phase domain wall systems [46–50]. The reason for observing the splitting in LEED is the presence of a large number of anti-phase domains within the incident electron beam spot size on the sample which leads to the interference between the scattering vectors of the overlayer domains [45, 50]. For example, in a kinematical calculation study, Zeppenfeld et al. show that in any domain wall system of a $\sqrt{3} \times \sqrt{3}$ R30° superstructure, the first and second-order spots along the $\overline{\Gamma M}$ direction splits and the amount of atomic relaxation within the walls only influences the intensity of the spots [46]. Similarly, we observe the splitting of the first order alloy spots along $<01\overline{1}>$ both in our LEED data as well as our simulation, while the spots of the alloy superstructure along the $<11\overline{2}>$ direction shows no sign of any spot splitting. The splitting of the second-order spots along the $<01\overline{1}>$ is not easy to observe because of the high contrast at this point, which could be due to the additional moiré modulations. Also due to the tessellation, we were not able to resolve a complete moiré cell in the STM Data. The average experimental lattice size of the alloy superstructure discussed in the result section is very close to the one measured for the bulk tetragonal crystal of the DyAg$_2$ alloy (~5.23 Å), using x-ray and neutron diffraction methods [51]. This is not the case for the GdAu$_2$, which has nearly a similar bulk alloy lattice constant [51]. Considering the almost identical lattice constants of silver and gold crystals, we see different expansion ratios of the alloy layer. Thus, we can argue that the DyAu$_2$ alloy possibly reduces its surface energy by uniformly relaxing its in-plane lattice stress, i.e., by increasing the interatomic distance within the alloy layer. In contrast, the DyAg$_2$ structure reduces its surface free energy not only by a marginal increase of the interatomic distance of the periodic structure within the hexagonal tiles but by tessellation and the formation of domain walls between the hexagonal tiles [52]. This behavior points out to a possibly stronger interaction between the DyAg$_2$ surface alloy layer with the Ag(111) substrate. Despite the lack of uniform long-range atomic ordering, each hexagonal tile enclosed by stacking-fault domains walls (marked by rectangular yellow cells in Fig 4(d)) reveals a locally identical crystalline structure. A similar study on GdAg$_2$ surface alloy identifies local variation of the unit cell size (1 % strain) and change of orientation for each relative moiré pattern. This study identifies similar local formation of strained hexagonal tessellation upon low alloy formation temperature [27]. Varying the temperature of the substrate during sample preparation does not affect our results (see SI.6), which suggests that this structure is perhaps the only stable result that can be achieved for the DyAg$_2$ surface alloy on Ag(111). The surface alloy formation is hence governed by a delicate balance between the in-plane interatomic forces of the alloy layer and the electronic interaction i.e. bonding between the substrate and the top layer atoms. In other words, the atoms of the alloy layer favour settling on their energetically preferred adsorption sites to maintain a locally near-commensurate registry despite the resulting in-plane strain of the alloy layer. In the case of DyAu$_2$, interatomic bonds of the alloy layer endure this strain, so that a uniform structure with minimum ruptures is maintained. The DyCu$_2$/Cu(111) can be considered an intermediate case, where the interaction of the substrate with the alloy overlayer is more pronounced and in this particular case results in mirrored domains. In the case of the DyAg$_2$, the structure grows in its preferred configuration (within the tiles) until the accumulated strain is released by the formation of a domain wall. A more comprehensive understanding of the interplay between these forces in structure formation requires ab-initio calculation schemes that can identify the most stable structure(s) with the lowest cohesive energy.

## Conclusion

In conclusion, we investigated the atomic and mesoscopic structure of three Dy-based surface alloys formed on the fcc(111) noble metals Au, Cu, and Ag. The structure of each system is characterized using STM and LEED. We find a non-commensurate relation between the alloy and its respective substrate for all three samples. Each system exhibits a different kind of mesoscopic moiré structure suggesting a different behaviour of substrate/overlayer interaction. Both DyAu$_2$/Au(111) and DyCu$_2$/Cu(111) surface alloy systems show homogeneous moiré structures with non-identical valley contrasts in STM, but only the one of the former is aligned with the surface alloy unit cell, while for the latter it deviates substantially from the surface alloy orientation. In the case of the DyAg$_2$, we find a complex tessellation of the alloy overlayer consisting of anti-phase domains in form of hexagonal tiles that are surrounded by boundary walls, which locally break the general translation symmetry of the surface alloy layer. Crucially, the formation of anti-phase domains in this sample is the consequence of the in-plane strain within the DyAg$_2$ surface.

Overall, our study has demonstrated that the local atomic, as well as the mesoscopic structure of RE-NM surface alloys, are governed by a complex interplay between the intra-plane and inter-plane interactions of alloy and substrate due to lattice constant variation, mediated by strain. Lattice strain is justified by the significantly different atomic sizes of RE and NM atoms. Achieving active control of these competing processes offers great potential to design moiré patterns with complex spin-dependent band structures and hence to template nanostructures for site-specific absorption of (organic) adsorbate, or the formation of heteromolecular nanostructures [53].

## Acknowledgements

The experimental work was funded by the Deutsche Forschungsgemeinschaft (DFG, German Research Foundation) - TRR 173 - 268565370 Spin + X: spin in its collective environment (Projects A02). B.S. acknowledges financial support by the Dynamics and Topology Center funded by the State of Rhineland Palatinate. We thank Azadeh Kadkhodazadeh for providing us with the STM data used in SI.5.

# Atomic and mesoscopic structure of Dy-based surface alloys on noble metals

*Supplementary information*

Sina Mousavion[1], Ka Man Yu[1], Mahalingam Maniraj[1,2], Lu Lyu[1], Johannes Knippertz[1], Benjamin Stadtmüller[1,3] Martin Aeschlimann[1]

[1]Department of Physics and Research Center OPTIMAS, University of Kaiserslautern, Erwin-Schrödinger-Straße 46, 67663 Kaiserslautern, Germany
[2]Departent of Physics, Indian Institute of Technology Bombay, Powai, Mumbai 400076, Maharastra, India
[3]Institute of Physics, Johannes Gutenberg University Mainz, Staudingerweg 7, 55128 Mainz, Germany
*Email: sina.mousavion@gmail.com, bstadtmueller@physik.uni-kl.de

# 1. The thickness of the domain boundary walls in tessellated DyAg$_2$

The evaluation was performed in Gwyddion. 15×15 nm$^2$ STM image (Fig 3(c)) was FFT filtered to enhance the protrusions' contrast. Line scans were plotted between each hexagonal tile for all three directions within this scanned area. The maxima were extracted for each set of profiles. The average distance of maxima for all three directions is calculated.

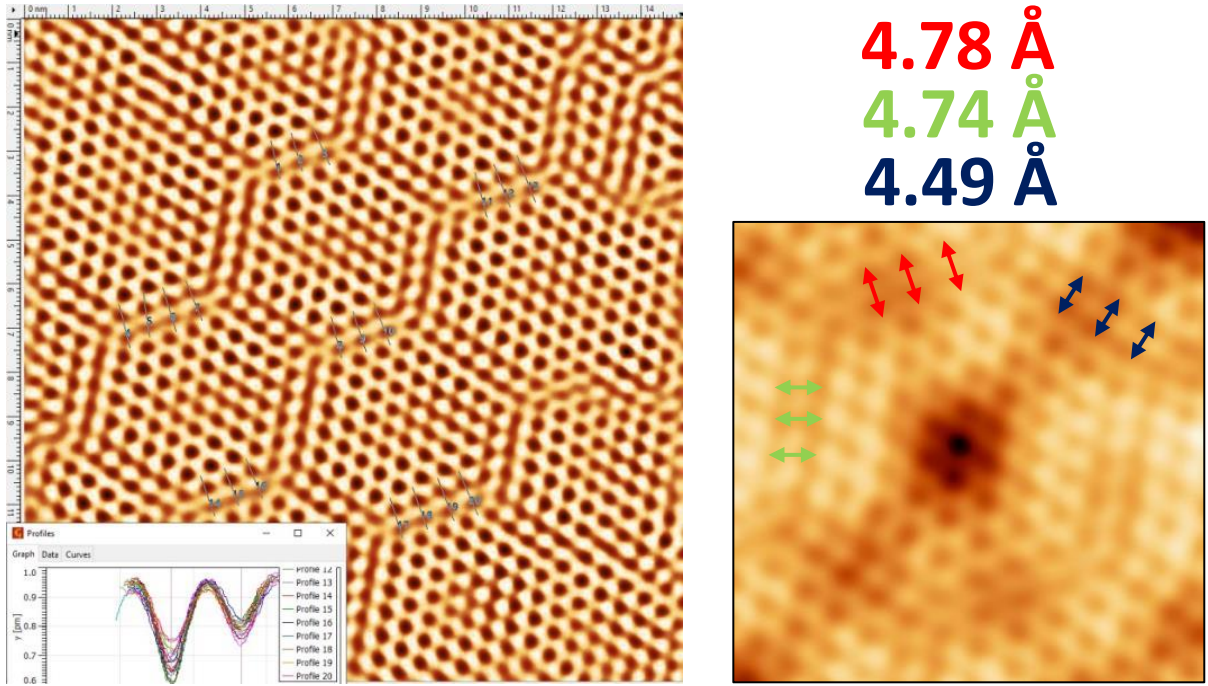

**Figure SI. 1 Left:** FFT-filtered image of the STM data presented in Fig 3 (c). The line profiles in the domain crossings in one direction are plotted in the inset. **Right:** average atomic distance of domain boundaries in three directions

# 2. The nearest neighbor distance map of the DyAg$_2$/Ag(111) superstructure

The evaluation was performed by creating a Euclidean distance map (EMD) for the 15×15 nm$^2$ STM image (Fig 3(c)) in ImageJ. EMD visualizes the distance between the black features. The scanned image was FFT filtered to enhance the black features. The image was recalibrated and converted to 8-bit for thresholding. The binary watershed was used for separating the features that might have been considered one. Using the article analyzer, the central position of each feature was extracted as a CSV file. Using MATLAB, a ball model is created according to the extracted positions and using identical circles as the black features. The EMD was then generated using this ball model in ImageJ. We can notice two things from Figure SI. 2. Firstly the larger distances inside the domain boundary are due to a deformed lattice (larger yellow patches). Secondly, the features within these boundary tiles are almost equidistantly distributed, indicating a highly crystalline structure with minimum strain.

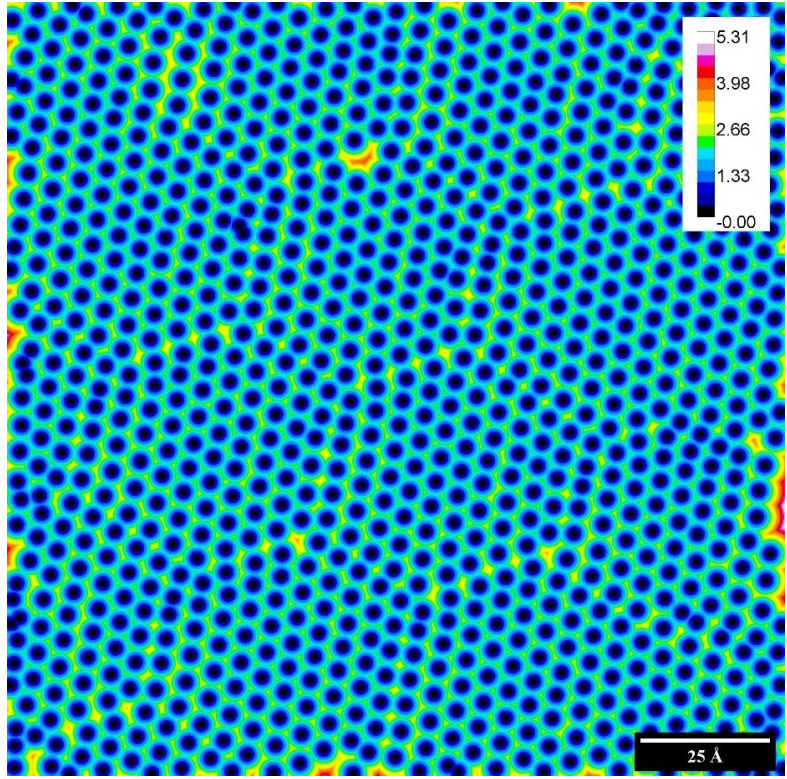

**Figure SI. 2** EDM of the $15 \times 15$ nm$^2$ STM scan from Fig 3(c). The colour bar is the nearest distance from a circular feature in angstrom

## 3. Structure factor calculation for uniform and tessellated DyAg$_2$/Ag(111) superstructure models

Two sets of ball models were generated in MATLAB according to the $\begin{bmatrix} 2.08 & 1.04 \\ -1.04 & 1.04 \end{bmatrix}$ superstructure matrix, identified in the LEED section. Fig SI. 3(a) is an ideal uniform structure and Fig SI. 3(b) is a tessellated model. The tessellated model is created by repeating a $5 \times 5$ tile using a $14.5 \times Ag_{sub}$ translation vector along the substrate's $<01\bar{1}>$ directions. This results in domain walls with a uniform width of 4.42 Å. The structure factor $S(K)$ for each alloy model was calculated as

$$S(K) = \sum_{n=0}^{N} f_k e^{-i(K \cdot r_n)} \qquad (1)$$

where $K = (k_x, k_y)$ is the scattering wave vector, N is the number of the atoms and $r_n$ is the position of the $n^{th}$ atom in the real space. The atomic form factor $f_k$ is considered one in our model, treating all atomic elements as equals [1]. The scattering intensity is calculated as

$$I(K) = \frac{1}{N^2} |S(K)|^2 \qquad (2)$$

In our calculations, only the atoms of the alloy layer are considered to clearly demonstrate the effect of an anti-phase arrangement on the splitting of the diffraction spots.

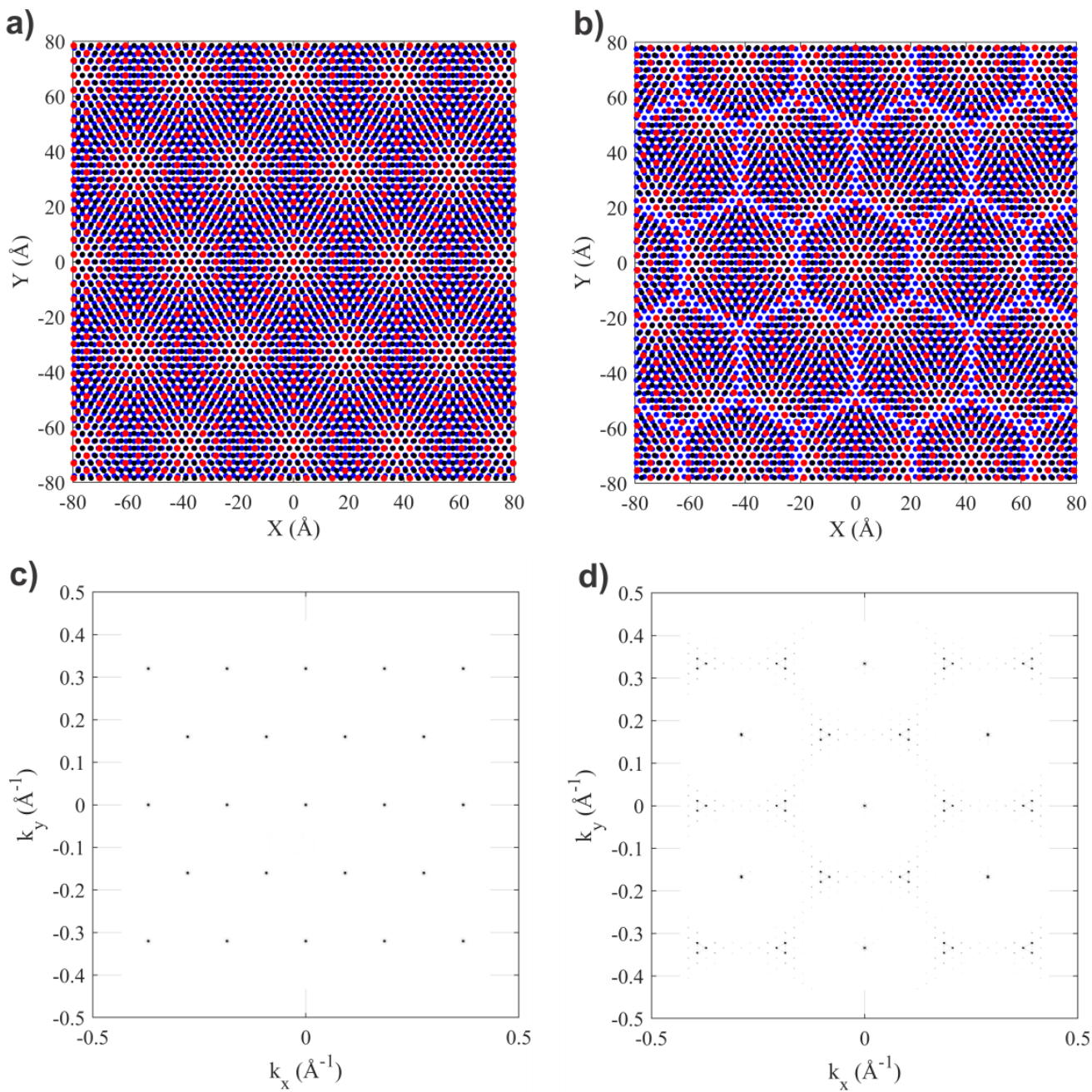

**Figure SI. 3 ball model of a uniform DyAg$_2$/Ag(111) structure (a) and a tessellated DyAg$_2$/Ag(111) structure (b) according to the LEED superstructure matrix. Ag atoms of the substrate, Dy atoms and Ag atoms of the alloy layer are depicted by blue, red and black spheres, respectively. Diffraction simulations of the uniform model (c) and the tessellated model (d)**

## 4. Moiré unit cell calculation

In this model, we only used the proposed surface alloy unit cells, evaluated from our LEED result. Using the following formulas. Calculations of the moiré unit cells' parameters, provided in Table 1. are based on the general model by Hermann [2] for a hexagonal substrate lattice ($a$) with isotropically scaled superstructure ($b$) which is rotated by ($\alpha$) with respect to the substrate. Moiré lattice (M) and the rotation relative to the substrate ($\gamma$) is calculated as

$$M = \frac{a}{\sqrt{1 + p^2 - 2p \cos(\alpha)}} \quad (3)$$

,

$$\gamma = \cos^{-1}\left(\frac{\cos \alpha - p}{\sqrt{1 + p^2 - 2p \cos(\alpha)}}\right) \quad (4)$$

where $p = a/b$, is the scaling factor. $a$ is the substrate lattice constant and $b$ is the estimated nearest neighbor distance inside the surface alloy unit cell. If we assume that the atomic units are equidistantly placed within the alloy layer, then $b = \frac{\sqrt{3}}{3} a_{alloy}$ and $\alpha$ is the surface alloy unit cell rotation value with respect to the substrate's <11$\bar{2}$> directions.

## 5. STM image of both mirrored domains of DyCu$_2$/Cu((111) sample

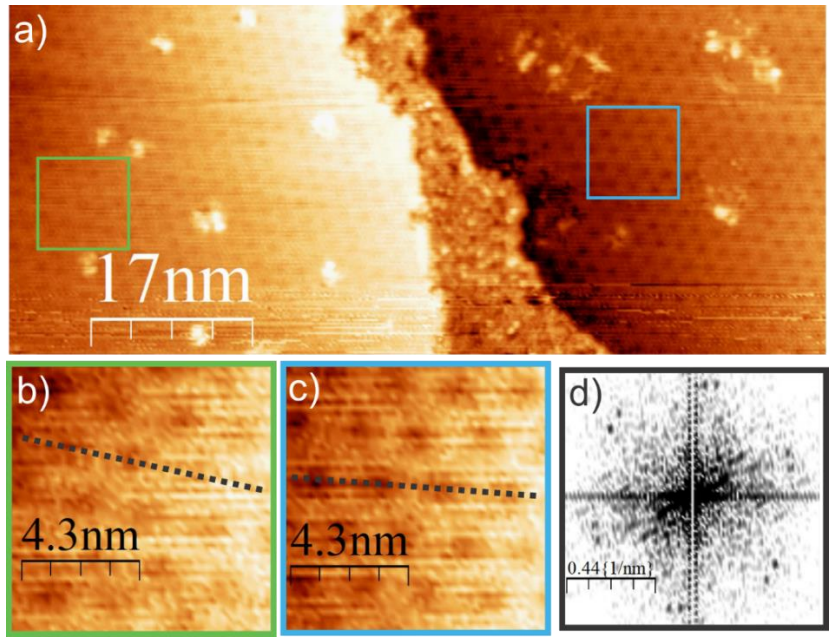

**Figure SI. 4** STM image of DyCu$_2$/Cu(111) sample (a) scan area with adjacent mirrored domains (b), (c) enlarged images of the green and blue squares respectively (d) FFT of (a). Measurement at constant current mode, $I_{set}$= 5.8 nA, $V_{Bias}$= 551 mV and at room temperature.

## 6. LEED results for DyAg$_2$/Ag(111) preparation at different substrate temperatures

Samples were carefully prepared at identical evaporation conditions (flux and time). For each sample, the substrate was cleaned using multiple cycles of sputter/annealing mentioned in the experimental section. Prior to the evaporation LEED confirmed that no superstructure was present on the surface. LEED images for the surface alloy samples were acquired after the sample reached room temperatures. For temperatures between 328 – 359 °C, almost no change is observed. 373 °C and 385 °C show a decrease in spot intensity but no observable change in the pattern. At 394 °C only the silver spots are visible, which suggests that the superstructure is possibly desorbed.

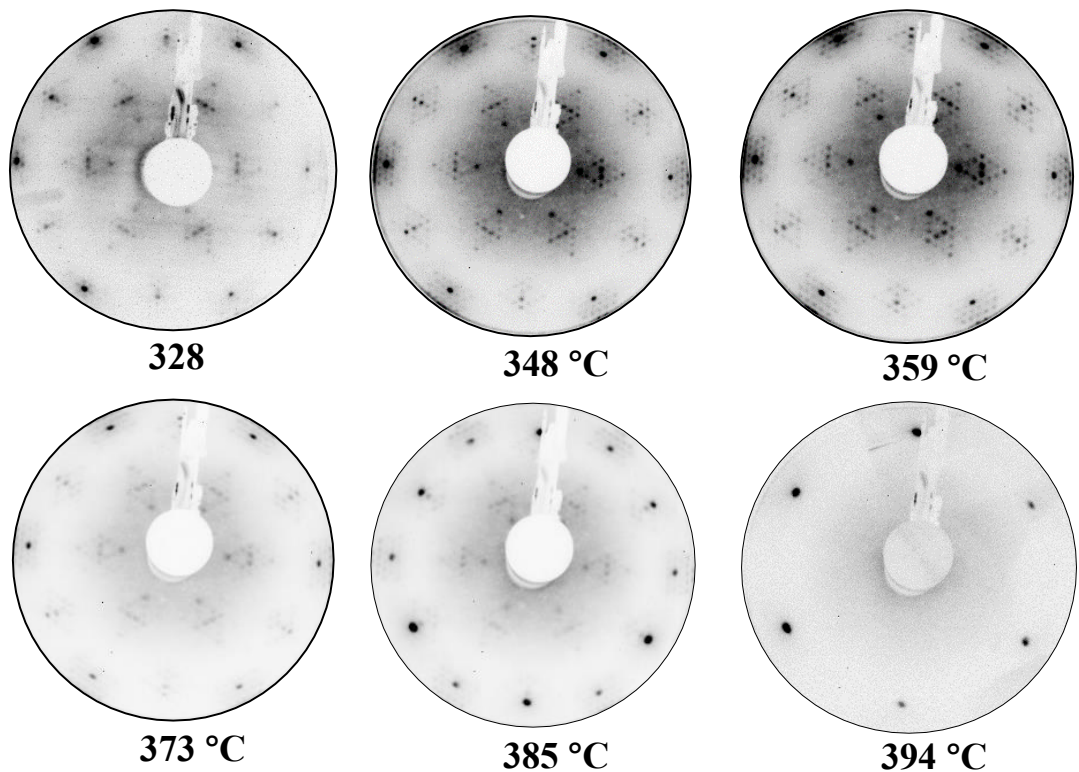

**Figure SI. 5 LEED images at E = 55 eV for DyAg$_2$/Ag(111) samples prepared at various substrate temperatures.**